\documentclass[12pt]{article}
\usepackage{amsmath,amssymb}
\begin{document}

\newcommand{\e}{\epsilon}
\newcommand{\beq}{\begin{equation}}
\newcommand{\eeq}[1]{\label{#1}\end{equation}}
\newcommand{\bea}{\begin{eqnarray}}
\newcommand{\eea}[1]{\label{#1}\end{eqnarray}}
\renewcommand{\Im}{{\rm Im}\,}
\renewcommand{\Re}{{\rm Re}\,}
\newcommand{\diag}{{\rm diag} \, }

\begin{titlepage}
\begin{center}

\vskip 4 cm

{\Large \bf Long String Dynamics in Pure Gravity on AdS$_3$ }

\vskip 1 cm

{Jihun Kim  and Massimo Porrati}

\vskip .75 cm

{\em Center for Cosmology and Particle Physics, \\Department of Physics, New York University, \\4 Washington Place, New York, NY 10003, USA}

\end{center}

\vskip 1.25 cm

\begin{abstract}
\noindent  We study the classical dynamics of a completion of pure AdS$_3$ gravity, whose
only degrees of freedom are boundary gravitons and long strings. We argue that the
best regime for describing pure gravity is that of ``heavy'' strings, for which  back-reaction  effects on the metric 
must be taken into  account. We show that once back-reaction is properly accounted for, regular finite-energy states 
are produced by heavy strings even  in the infinite-tension limit. Such a process
is similar to, but different from, nucleation of space out of a ``bubble of nothing."
\end{abstract}
\end{titlepage}
\newpage

\tableofcontents
\noindent\hrulefill
\bigskip



\section{Introduction and summary}
This paper is dedicated to Valery Rubakov in occasion of his 60th birthday. Valery has been a pioneer and a master in understanding the role of 
non-perturbative solutions to field equations in quantum field theory. This paper is devoted to a particular case of 
soliton dynamics. Though limited in scope, we believe that it contains some
results worth reporting. We hope that its readers will consider it also a worthy 
tribute to Valery's work.

Pure gravity in three dimensions does not propagate local degrees of freedom, as
a simple counting argument shows. six of the twelve Hamiltonian degrees of freedom of
the 3D graviton, $g_{\mu\nu}$, are removed by gauge invariances and the other 
ones are removed by 3+3 constraints that follow from Einstein's equations.
So, 3D gravity does not propagate gravitational waves. In the presence of a 
negative cosmological constant, pure gravity still exhibits a nontrivial dynamics, 
because there exist boundary 
gravitons~\cite{bh} and black hole solutions~\cite{btz}. 
The Einstein-Hilbert
action of  pure gravity with negative cosmological constant, $-1/l^2$,  is
\beq
S_{EH}={1\over 16\pi G}\int d^3x \sqrt{-g}\left ( R + {2\over l^2}\right).
\eeq{m00}  

Boundary gravitons exist because the asymptotic metric of 3D Anti de Sitter space (AdS) is preserved by
a set of diffeomorphisms that act non-trivially on the boundary. 
Specifically, the condition of being asymptotically AdS$_3$ means that the metric
has the form~\cite{bh}
\beq
g_{tt}= -r^2/l^2 + O(1), \qquad g_{t\phi}=O(1) , \qquad g_{tr}=O(r^{-3}),
\nonumber 
\eeq{m0}
\beq
g_{rr}=l^2/r^2 + O(r^{-4}), \qquad g_{r\phi}=O(r^{-3}), \qquad
g_{\phi\phi}=r^2 + O(1).
\eeq{m1}
These boundary conditions are preserved by diffeomorphisms with asymptotic 
form 
\bea
\zeta^t&=&l[f(x^+) + g(x^-)] + {l^3\over 2r^2} [\partial^2_+ f^+(x^+) + 
\partial^2_-g(x^-)] + O(r^{-4}), \nonumber \\
\zeta^\phi&=&[f(x^+) - g(x^-)] - {l^2\over 2r^2} [\partial^2_+ f^+(x^+) - 
\partial^2_-g(x^-)] + O(r^{-4}), \nonumber \\
\zeta^r&=& -r[\partial_+f(x^+) + \partial_-g(x^-)] + O(r^{-1}).
\eea{m3}
The allowed diffeomorphisms are parametrized by two arbitrary functions $f(x^+)$, $g(x^-)$, 
each depending on only one of the 
two boundary light cone coordinates $(x^\pm= t/l\pm \phi)$. The time $t$
and the angular coordinate $\phi\sim \phi + 2\pi$ parametrize the AdS$_3$
boundary, while $r$ is its radial coordinate. The boundary is at $r=\infty$ and
$2\partial_\pm=l\partial/\partial t \pm \partial/\partial \phi$. 

The classical Poisson brackets associated to the asymptotic 
diffeomorphisms~(\ref{m3}) define two Virasoro algebras with equal 
central charge 
$c=3l/2G$~\cite{bh}; therefore, after quantization, the Hilbert space of any quantum 
gravity with the same asymptotics --whether pure or with matter-- must fall 
into unitary representations of the Virasoro algebras. This purely 
kinematical fact
has a deep consequence if one further assumes  that quantum gravity on AdS$_3$
is dual to a 2D conformal field theory (CFT)~\cite{mal}. 
Modular invariance of the CFT, discreteness of the spectrum 
and the existence of an $Sl(2,C)$ invariant state
with conformal weights $\Delta=\bar{\Delta}=0$ then
 imply that the asymptotic density of states at levels 
$(\Delta,\bar{\Delta})$ is~\cite{cardy}
\beq
d(\Delta,\bar{\Delta})\equiv e^S= \exp\left(2\pi \sqrt{c\Delta/6} + 
2\pi\sqrt{c\bar{\Delta}/6}\right).
\eeq{m4}

Rotating black hole solutions for pure 3D AdS gravity~(\ref{m1}) do 
exist~\cite{btz}. Their metric depends on two parameters: mass $M$ and angular
momentum $J$. The metric is~\cite{btz}
\bea
ds^2 = - N^2 dt^2 + N^{-2} dr^2 + r^2 ( N^{\phi} dt + d {\phi})^2 ,\nonumber \\
N^2 = - 8GM + \frac{r^2}{l^2} + \frac{16 G^2 J^2}{r^2} , \qquad N^{\phi} = 
\frac{4GJ }{r^2} .
\eea{m5}
After the identification $\Delta+c/24=(Ml+J)/2$, 
$\bar{\Delta}+c/24= (Ml-J)/2$, the Cardy formula~(\ref{m4}) matches the 
Bekenstein-Hawking formula for the entropy of rotating black holes~\cite{strom}
\beq
S=S_{BH}=2\pi r_h/4G, \qquad r_h=l\sqrt{4GM +4G\sqrt{M^2-J^2/l^2}}.
\eeq{m6}

The result of ref.~\cite{strom} is general. In particular, it does not depend 
on the matter content of the AdS$_3$ bulk theory. Amusingly, pure gravity
seems to defy general formulas~(\ref{m4},\ref{m6}). Indeed, as noticed 
in~\cite{chvd}, the asymptotic dynamics of eq.~(\ref{m00}) is described by a 
Liouville action. Upon quantization, Liouville theory becomes an unusual 
conformal field theory because of two features. The first is that its spectrum 
does not include an $Sl(2,C)$ invariant state. Physical 
states obey instead
the ``Seiberg bound''~\cite{sei} $\Delta,\bar{\Delta} > (c-1)/24$. The second
is that physical states are only plane-wave normalizable, because the spectrum
of Liouville theory is continuous. These properties are well established in 
consistent quantizations of Liouville theory at $c>1$~\cite{tesch}.

The reduction of pure gravity to a boundary Liouville theory is most easily proven by writing the Einstein-Hilbert  
action~(\ref{m00}) in terms of two $Sl(2,R)$ Chern-Simons theories~\cite{town} 
\beq
S_{EH}=S_{CS,k}[{A}] -  S_{CS,k}[\tilde{A}], \qquad k=l/4G.
\eeq{m7}
Denoting by $t^a$ the three $Sl(2,R)$ generators in the fundamental representation, the Chern-Simons action 
is
\beq
S_{CS,k}[A] =  {k \over 4\pi}\mathrm{Tr}\,\int_M  \left (A\wedge d A + {2
\over 3} A \wedge A \wedge A \right) + \mbox{ boundary terms}.
\eeq{m8}
The gauge potentials $A,\tilde{A}$ are related to the dreibein $e^a$ and spin connection $\omega^a$ by
\beq 
A^a= \omega^a+ {e^a \over l} , \qquad \tilde{A}^a = \omega^a -{e^a \over l} , \qquad A=A^at^a.
\eeq{m9}
Some of the equations of motion derived from~(\ref{m7}) are constraints. In the gauge $A_-=\tilde{A}_+=0$
{\em when the 3D space is topologically the product of a 2D disc $D$ and the real line $R$},  they 
imply $A_r=U^{-1}d U$, $\tilde{A}= V^{-1} d V$, with $U$ ($V$) an $Sl(2,R)$ valued function of $r,x^+$ ($r,x^-$). 
Substituting the solution of the constraints into the Chern-Simons action, bulk terms disappear and the action reduces to a boundary term. 
This term is the 2D chiral Wess-Zumino action~\cite{w-j,emss}. Further constraints,
following from the requirement that $A,\tilde{A}$  give an asymptotically AdS metric, reduce the Wess-Zumino action to 
a Liouville action~\cite{chvd}.

An attentive reader should have noticed an unwarranted assumption here. 
We {\em assumed} that the 3D space was 
topologically global AdS$_3$ to arrive at a Liouville action. In the presence of black holes, i.e. horizons, or of time-like
singularities associated with point-like particles in the bulk, the action at the $r=\infty$ boundary must be supplemented with 
other terms at the inner boundary/horizon. A possible interpretation of these terms is that they describe the states of the AdS$_3$ quantum gravity; more precisely the primary states in each irreducible representation (irrep) of the 
$Virasoro \times Virasoro$ algebra acting on
the Hilbert space of quantum AdS$_3$ gravity~\footnote{Ref.~\cite{emss} uses
the ``constrain first'' Hamiltonian formalism to study 
the effect
of point-like insertions and nontrivial topology for compact-group Chern-Simons theories.}. The role of the boundary Liouville would be then simply to describe, in 
each irrep, the Virasoro descendants (cfr.~\cite{mart}). In this interpretation, other information is needed to determine the spectrum of primary operators. 

One hint that pure gravity could nevertheless have the same spectrum of primaries as Liouville theory comes from canonical quantization of pure gravity. Already in the 1990's, it was shown that the the wave functions obtained by quantization of $Sl(2,R)$ Chern-Simons theory are Virasoro conformal blocks~\cite{ver}. Two $Sl(2,R)$ Chern-Simons 
actions are combined into the action of pure gravity so the Hilbert space of pure gravity must be (a subspace of) 
the product of each Chern-Simons Hilbert space. In a forthcoming publication we will argue that the pure gravity 
Hilbert space is the target space of conformal field theories {\em with continuous spectrum and obeying the 
Seiberg bound}~\cite{kimporrati} (cfr.~\cite{w3d}).  Assuming from now on that this result holds, we conclude that 
 pure gravity in AdS$_3$ should contain states that can reach the boundary at a finite cost in energy, since states confined to the interior of the AdS space have a discrete spectrum. So one natural question to ask is, what are those states?

The mass of such states must be large in AdS units: $Ml \gg 1$, otherwise gravity could not be called ``pure" in any sense. The states cannot be massive particles, which cannot reach the AdS boundary. Indeed, there is
only a natural candidate for such states: they must be long strings. These states were already invoked as a possible 
solution to certain problems of the partition function of Euclidean pure gravity in~\cite{mawit}.

The rest of this paper is devoted to studying the effect of long strings in AdS$_3$ gravity. 
Section 2 will summarize known features of long strings in the probe approximation, which holds 
when back-reaction on the metric and quantum string dynamics effect can both be neglected. This
happens when the string tension $T$ is in the range $l^{-2} \ll T \ll G^{-1}l^{-1}$. 

Section 3 describes the case of ``light'' strings, which were studied in details in~\cite{gkrs}: 
$T\lesssim l^{-2}$. It is a regime where back-reaction can be neglected but quantum effects 
cannot. This is an interesting case, but far from pure gravity, as we will argue using some results 
of ref.~\cite{gkrs}.
   
Section 4 studies the ``heavy'' string case, $T\gtrsim G^{-1}l^{-1}$, when back-reaction cannot be
neglected. We argue that this regime is the best suited to describe a pure gravity 
theory containing BTZ black holes and no state below the
Seiberg bound. We further show that in order to recover the mass gap predicted by the 
Seiberg bound the string tension
must be Planckian $T=O(G^{-2})\gg G^{-1}l^{-1}$. This is the limit 
$T\rightarrow \infty$, which is nonsingular thanks to back-reaction effects. Finite 
mass BTZ states arise through a process similar to nucleation of the universe out of a
``bubble of nothing.''~\footnote{Differently from the quantum nucleation case, the process under consideration here is a classical one, in which the initial state contains a long string approaching the boundary in the far past. This is good,
since up-tunnelling from a bubble of nothing~\cite{w-bn} 
is forbidden in AdS space~\cite{ab}.}

\section{Long Strings in Probe Approximation}
If short string dynamics and back-reaction are negligible, as it happens when the string tension is in the intermediate range 
$l^{-2} \ll T \ll G^{-1}l^{-1}$, the effects of long strings can be described in the 
probe approximation. The long string probe is located at radial position 
$r=R(\phi)$ and its classical action is made of two terms~\cite{sw}. 
One is proportional to the
area spanned by the string world-sheet $\Sigma$, the other is proportional 
to the volume enclosed by the world-sheet.
\beq
S= TA(\Sigma) - \frac{q}{l^3} V(\Sigma).
\eeq{m10}
The second term requires coupling the string to an antisymmetric two form. The
world-sheet action of the string  thus acquires a term
\beq
S=... + q\int_\Sigma dX^\mu \wedge dX^\nu B_{\mu\nu}.
\eeq{m11}
The two-form $B$ 
is analogous to the Kalb-Ramond form of fundamental strings. It 
possesses the gauge invariance $B \rightarrow B + d\Lambda$ and its
bulk action is
\beq
S_{B}=-(1/12)\int_{AdS_3} H\wedge * H, \qquad H=dB.
\eeq{m12}
This action does not propagate
any degree of freedom in 3D. So, the bulk theory in the presence of the form 
$B$ is still pure gravity, but with a cosmological constant  that depends on
the value of the field strength $H$. The field strength  is quantized in units of $q$, the
two-form charge of the string~\cite{bt,bp}. 

The asymptotic value of the string action~(\ref{m10}) is best written in terms of a redefined radial coordinate
$\varphi$, the induced world-sheet metric $h$, and the world-sheet scalar curvature $R$~\cite{sw} as
\bea
S&=& {Tl^2} \int_\Sigma d^2\sigma 
\sqrt{h} \left[ (1-q)e^{2\varphi}/4 + (\partial \varphi)^2/2 + \varphi R/2  - R/4 + 
O(e^{-2\varphi})\right], \label{m13} \\
r/l &=& e^\varphi + e^{-\varphi} \varphi R l^2 + O(e^{-2\varphi}).
\eea{m14} 
To reach the boundary with finite energy, one must set $q=1$. At $q=1$~(\ref{m13}) becomes  the Liouville action. Its central charge is $c_L=1+12\pi Tl^2$. Quantum effects can be
neglected in the semiclassical regime for Liouville theory, that is when $c_L \gg 1$, hence when $T\gg l^{-2}$.
Crossing the brane, the cosmological constant changes and so does the central charge $c=3l/2G$. If we call
$l_+$ the AdS radius outside the brane and $l_-$ the radius inside, the central-charge change is
\beq
3l_+/2G - 3l_-/2G \equiv \Delta c= c_L\approx 12\pi Tl^2.
\eeq{m15}
Back-reaction effect can be neglected when $\Delta c/c \ll 1$, hence when $T\ll G^{-1}l^{-1}$. This inequality on the
other hand implies that the energy gap between the vacuum and the long string states, given by the Seiber bound
with $c=c_L$, is $E=(c_L-1)/12= \pi Tl^2 \ll (c-1)/12$. So, the theory contains states with energy well
below the BTZ black hole threeshold. It is therefore doubtful whether we can call gravity plus strings in the
regime $l^{-2} \ll T \ll G^{-1}l^{-1}$ ``pure.'' 

The most obvious method for increasing the gap is to make $T\gtrsim G^{-1}l^{-1}$ and take full 
account of the back-reaction. This will be done in section 4. In the next section we examine a more exotic 
possibility. Namely, we study the dynamics of light strings with tension $T\ll l^{-2}$. Though a theory with strings of
tension smaller than the AdS scale contain a large number of light states, maybe it could still bear resemblance
to pure gravity if these states decouple in the limit that the string coupling constant goes to zero. In
next section we will use the results of~\cite{gkrs} to argue against this possibility.

\section{Light Strings and the Absence of BTZ States}
Strings in AdS$_3$ with background NS forms can be studied to all orders in 
$\alpha'=l_s^2=1/2\pi T$. One can find in particular exact expressions for the
generators of the target space Virasoro algebras~\cite{gks}. The low-tension
region $l_s \gtrsim l$ may seem quite the opposite of pure gravity, since it 
contains an abundance of light degrees of freedom. One exotic possibility 
is to decouple all the unwanted states by sending the string coupling 
constant  $g_s$ to zero. Since $g_s^2= G/l_s$~\cite{gks}, decoupling means 
that we are sending $l_s\rightarrow \infty$ while keeping $G$, $l$  finite and
$l/G\gg 1$. The last condition guarantees that the AdS space is still 
macroscopic  and concepts such as black hole, metric etc. are meaningful. 
The first condition {\em may} decouple stringy excitations leaving only BTZ
states.  

To check if decoupling is actually possible we must parametrize our theory 
in terms of quantities that remains valid beyond the point-particle limit. 
So, instead of the ratio $l/l_s$ we should use the level $k$ of the
$Sl(2,R)$ world-sheet current algebra and  use the target space central charge
$c$ instead of $l/G$. In the semiclassical, point particle limit, $k= l^2/l_s^2$
and $c=3l/2G$. 

Ref.~\cite{gkrs} argues that a sharp phase transition occurs at $k=1$. For
$k>1$, the asymptotic density of states at high energy is dominated by BTZ
black holes and the target space theory has an $Sl(2,R)\times Sl(2,R)$ 
invariant  vacuum. For $k<1$ neither the  vacuum nor the BTZ black hole states are normalizable. 
The asymptotic density of states is dominated 
by weakly-coupled long strings. The first property agrees with 
expectations from canonical quantization of pure gravity and its similarities 
with Liouville theory. The second property seems to contradict the fact that
BTZ black holes are the only primary states in pure gravity. Nevertheless, it could be that 
at $k<1$ weakly-coupled long strings are just BTZ states in disguise.

The last possibility seems unlikely and in any case a better argument exists 
against the decoupling limit. The problem arises because in a  conformal field 
theory where the lowest conformal weight of a physical primary operator 
is not $\Delta=0$, but some $\Delta_m>0$, the effective central charge
appearing in Cardy's formula~(\ref{m4}) is 
$c_{eff}= c-24\Delta_m$~\cite{cardy}. The Seiberg
bound $\Delta\geq (c-1)/24$ then tells us that in ``Liouville like'' pure
gravity $c_{eff}=1$. 

On the other hand, ref.~\cite{gkrs} found that the effective
target space central charge for the long string gas is~\footnote{
Ref.~\cite{gkrs} studies type II and heterotic superstings. Formulae for the
bosonic string are the same as those for the right-moving sector of 
the heterotic string.} 
\beq
c_{eff}= \left\{ \begin{array}{ll} 
6g_s^{-2} (2-1/k) & \mbox{for type II superstings} \\ 6g_s^{-2} 
(4-1/k) & \mbox{for bosonic strings} \end{array} .\right. 
\eeq{m16}
Setting $c_{eff}=1$ we have
\beq
k= \left\{ \begin{array}{ll} 
1/2 + O(g^2) &\mbox{for type II superstings} \\ 1/4+ O(g^2) 
&\mbox{for bosonic strings} \end{array} .\right. 
\eeq{m17}
On the other hand, the target space central charge is~\cite{gkrs}
\beq
c= \left\{ \begin{array}{ll} 
6g_s^{-2} k &\mbox{for type II superstings} \\ 6g_s^{-2} (k+2) &\mbox{for 
bosonic strings} \end{array} .\right. 
\eeq{m18}
So, in both cases $g_s\rightarrow 0$ implies $c\rightarrow \infty$, while we 
want to keep $c\gg 1$ but {\em finite}. 

If we had tried to keep $c$ finite in the limit $g_s\rightarrow 0$ we would 
have also run into a contradiction because $c_{eff}$ would have become either
negative or larger than $c$. Both these possibilities are forbidden in unitary CFTs. 

\section{Heavy Long Strings}

In the regime $T \gtrsim G^{-1} l^{-1}$ the metric is deformed by the back-reaction of the string.
The process that can lead to formation of massive point particles or BTZ black hole is the collapse of a long string 
arbitrarily close to the boundary of AdS$_3$ in the far past. 
This is what we examine at the classical level in this section.
The  collapse of shells of matter with various equations of state was considered in~\cite{mo}.

Consider first the collapse of a shell of matter  with rotational symmetry --which is a closed string in two 
space dimensions-- arriving from a radial position arbitrarily close to  the boundary of an
asymptotically AdS space in the far past.  Inside the shell the metric is pure AdS$_3$ and outside is a non-rotating BTZ black hole. 
The metrics inside and outside a shell with world-sheet $\Sigma$ are
\bea
ds^2_{-} &=& -\left( 1 + \frac{r^2_-}{l^2_-} \right) dt^2_-  + {\left(1 + \frac{r^2_-}{l^2_-} \right)}^{-1} dr^2_- + d {\phi}^2_-  ,\nonumber \\
ds^2_{+}&=& -\left( -8GM + \frac{r^2_+}{l^2_+}\right) dt^2_+  + {\left( -8GM + \frac{r^2_+}{l^2_+}\right)}^{-1} dr^2_+ + d {\phi}^2_+ .
\eea{met}
Here the subscript $(-)$ is used for variables defined inside the shell and $(+)$ for those outside it.
If the string has no angular motion, we can define ``proper time'' by moving on the world-sheet at fixed $\phi$ and parametrize $\Sigma$ as
\beq
ds^2_{\Sigma} = -d {\tau}^2  + \left(  R(\tau) \right)^2 d {\phi}^2 ,
\eeq{ws}
where $R(\tau)$ is the radius of the string.

The discontinuity of the extrinsic curvature $K^{\pm}_{ij}$  across the shell is related to the stress-energy tensor of the string $S_{ij}$ on $\Sigma$ by the so-called Israel boundary conditions; precisely
\beq
 \gamma^{+}_{ij} - \gamma^{-}_{ij} = 8 \pi G S_{ij}, \qquad \gamma^{\pm}_{ij} = K^{\pm}_{ij} - g_{ij}K^{\pm} .
 \eeq{disc}
 It is convenient to study the string dynamics using a comoving frame, spanned by proper time and 
 $(1/R)\partial_{\phi}$. In such frame $S_{ij} = \diag(T,-T)$, 
and the discontinuity in $\gamma^{\pm}_{ij}$, which we call $\gamma_{ij}$, is
\beq
\gamma_{\tau \tau } = - \frac{1}{R} (  \beta_+ - \beta_- ) , \qquad \gamma_{\theta \theta} = \frac{d}{d R} (\beta_+ - \beta_- ).
\eeq{gamma}
Here $\beta_{\pm}= \sqrt{ {\dot{R}}^2 - 8GM_{\pm} + {R^2}/{l^2_{\pm}}}$, while $R(\tau)$ is the position of the string and $M_- = -1/8G$.

Although we can easily solve exactly the single  equation obtained from~(\ref{disc},\ref{gamma}),
examining its asymptotic behavior is sufficient for our purpose. 
\bea
8 \pi G T R &=& \sqrt{ {\dot{R}}^2 + 1 + \frac{R^2}{l^2_-} } - \sqrt{ {\dot{R}}^2 - 8GM + \frac{R^2}{l^2_+} } \nonumber \\
&=& \left( \frac{1}{l_-} - \frac{1}{l_+} \right) R + \frac{1}{2R} \left[ ( l_- + 8GMl_+ ) + {\dot{R}}^2 ( l_- - l_+ ) \right] + {O}(R^{-3}).
\eea{asymp}

The leading order term tells us that the string tension should be 
\beq
8\pi GT = \frac{1}{l_-} - \frac{1}{l_+}.
\eeq{tension}
If the tension differs from this value, the string either cannot reach the boundary or reaches 
it with infinite radial speed. Eq.~(\ref{tension}) is the generalization to heavy strings of  
condition $q=1$  in section 2. 
We call a $T$ obeying eq.~(\ref{tension}) \emph{critical tension} and the string with such tension 
a \emph{critical string}.

For a critical string to exist, we must have $l_+ > l_-$, since $T$ is positive.
Then the subleading order term in asymptotic behavior~(\ref{asymp}) gives us an interesting bound on the mass 
of the collapsing string:
\beq
8GM \geq -\frac{1}{1 + 8 \pi GT l_+}.
\eeq{bound}

We can understand this mass bound better by expressing it in AdS unit.
At this point we have two length scale, $l_+$ and $l_-$, both of which can be used to convert energies into
dimensionless quantities. The radius of the asymptotically AdS metric outside the shell is $l_+$,
so, if we want to relate bulk energies to CFT weights, we have to use $l_+$ as our unit of lenght.

To compare with CFT and with section 2 it is convenient to redefine the AdS$_3$  energy as $E'=E+1/8G$. The  vacuum energy then vanishes,  all masses are positive and the mass bound becomes 
\bea
M' l_+ &\geq& \frac{l_+}{8G} \left( \frac{ 8 \pi G T l_+}{ 1 +  8 \pi G T l_+} \right)  \nonumber \\
&=& \frac{c_+}{12} \left( \frac{ 8 \pi G T l_+}{ 1 +  8 \pi G T l_+} \right).
\eea{adsbound}

The mass bound approaches zero as we send $8GTl_+$ to zero, so the tensionless limit cannot be 
related with the boundary Liouville theory obtained in section 2.

If  $8GM'l_+\gg 1$, on the other hand, we have finite mass gap
\beq
M 'l_+ \gtrsim \frac{c_+}{12},
\eeq{gap}
where $c_+ = 3l_+ / 2G$.
This agrees with the Seiberg bound in a theory with a large central charge $ c \gg 1$, as it is needed 
for classical geometry to make sense.

If we insists that this mass bound equals the Seiberg bound, we find $8 \pi G T l _+ = c_+ -1$.
This implies that the tension is  order of unity in Planck unit: $TG^2 \sim 1$.

We also have $l_- \sim G$ from critical tension condition~(\ref{tension}).
Therefore we are considering a process where a long string with large tension is 
nucleated at the boundary from an AdS$_3$ with Planckian curvature. Though similar to 
nucleation of an AdS ``bubble of nothing''~\cite{ab,w-bn}, the process is different. It is not a quantum
transition but a classical process: the collapse of a long string located at the boundary 
at past infinity. It is only thanks to back-reaction effects that the two ``infinities'' 
involved in the process, $1/l_-\sim 1/G$ and $T\sim 1/G^2$ cancel to give a finite result.

So, for $TG^2 \sim 1$, long strings can produce the right mass spectrum, consisting only of
BTZ black holes. Moreover, the large tension ensures that no unwanted low-energy states are being
added to ``pure'' gravity. 

We argued that long strings could account for BTZ black holes, but our attention was 
limited to non-rotating ones.
We want to conclude this section with some comment on the rotating BTZ case.
Inspired by previous consideration, it is  tempting to use long strings to explain rotating BTZ 
black holes through the collapse of a shell  formed by a rotating long string. We shall show next that this is impossible, 
as long as the world-sheet stress energy tensor $S_{ij}$ is diagonal.
The simplest case to analyze is a rotating BTZ with a string rotating at 
constant angular velocity and fixed radius $R$. This case suffices to show the general problem that one
encounters even in a more general setting. 

Suppose that inside the shell we have pure AdS$_3$ as before, but that the outside metric is
\bea
ds^2_{+} = - N^2 dt^2_+ + N^{-2} dr_+^2 + r_+^2 ( N^{\phi} dt_+ + d {\phi}_+ )^2, \nonumber \\
N^2 = - 8GM + \frac{r^2_+}{l_+^2} + \frac{16 G^2 J^2}{r^2_+} , \qquad N^{\phi} = 
\frac{4GJ }{r^2_+} .
\eea{rotating}
Notice that $ds^2_-$ is diagonal while $ds^2_+$ is not.
To compute $\gamma_{ij}$, however, the induced metrics on the world-sheet $\Sigma$ must be 
the same, i.e. $(ds^2_-) \vert_{\Sigma} = (ds^2_+)\vert_{\Sigma}$.
One way to accomplish this is by using a coordinate system spanned by
\bea
(e^-)_{\tau} =(\dot{t}_- , 0 , 0 ) ,\qquad (e^-)_{\theta} = ( 0 , 0 , 1/R) ,\nonumber \\
(e^+)_{\tau} =(\dot{t}_+ , 0 , \dot{\phi}_+ ) ,\qquad (e^+)_{\theta} = ( 0 , 0 , 1/R),
\eea{frame}
where $R$ is the radius of the string and $\dot{x}$ denotes the derivative of 
$x$ with respect to the proper time $\tau$.

This means that outside the world-sheet we use a rotating frame with constant angular 
velocity $\omega = \dot{\phi}_+ / \dot{t}_+$, which, in general, may be different from that of 
the string.
Both bases given in eq.~(\ref{frame}) are orthonormal if
\beq
\dot{t}_- = \frac{1}{\tilde{\beta}_-} , \qquad \dot{t}_+ = \frac{1}{\tilde{\beta}_+} , \qquad \omega = \frac{4GJ}{R^2}.
\eeq{ortho}
Here $\tilde{\beta}_{\pm} = \sqrt{ -8GM_{\pm} + R^2 / l^2_{\pm} + 16 G^2 J^2_{\pm} / R^2}$ 
with $M_- = -1/8G $, $J_-  = 0$ and $J_+ = J$.
In these coordinates systems, one finds that the extrinsic curvatures are
\beq
K^{\pm}_{\tau \tau } = \frac{d}{dR} \tilde{\beta}_{\pm} , \qquad K^{\pm}_{\tau \theta} = - 
\frac{4 G J_{\pm} }{R^2} , \qquad K^{\pm}_{\theta \theta} = - \frac{1}{R} \tilde{\beta}_{\pm}.
\eeq{ext}
Notice that these equations give $\gamma_{\tau \theta} = - 4GJ/R^2 $.
Since the origin of this term is the angular momentum of the BTZ black hole, 
we can not make it vanish by giving radial dynamics to the string.
As long as we consider physical configuration with angular symmetry, 
radial dynamics and rotations are the only motions we can introduce at the classical level.
This suggests, therefore, that we have to relax the string equation of state 
$\rho  = -p = T$ to explain rotating BTZ state. In fact, when the equation of state is $p=-\rho$ the string 
stress-energy tensor  $S^i_j$ is diagonal in {\em any} coordinate frame, whether rotating or not. 

One way to set $p\neq -\rho$ is by exciting degrees of freedom on the string. One such degree of 
freedom, the radial coordinate $R$, is always present, but other may exist as they do in fundamental
strings. 

One amusing agreement between long strings with equation of state $p=-\rho$ and Liouville theory
is that the latter contains only primaries with equal left and right conformal weights 
$\Delta=\bar{\Delta}$~\cite{tesch}. Since BTZ states must be primaries of the would be CFT dual,
such equality implies the vanishing of the BTZ angular momentum. 
 
At this point the  relation between long strings and rotating black holes is still unclear.
It is possible that we would need a completely different description for states giving rise to rotating black holes by gravitational collapse. 
However, nothing so far seems to forbid excited strings to produce rotating BTZ black holes. In
any case, production of BTZ black holes by long string collapse already showed intriguing features
and it is well worth more study.

\section*{Acknowledgements}
J-H.K. is supported by an NYU JAGA fellowship. 
M.P. is supported in part by NSF grant PHY-1316452. M.P also thanks the ERC Advanced Investigator Grant No.\@ 226455 {\em Supersymmetry, Quantum Gravity and Gauge Fields (Superfields)}.


\end{document}